\documentclass[twocolumn, switch]{article} 

\usepackage{preprint}

\usepackage{amsmath, amsthm, amssymb, amsfonts}

\usepackage[numbers,square]{natbib}

\usepackage[utf8]{inputenc}	
\usepackage[T1]{fontenc}	
\usepackage{booktabs} 		
\usepackage{nicefrac}		
\usepackage{microtype}		
\usepackage{lineno}		
\usepackage{float}			

\usepackage{lipsum}		

\usepackage{newfloat}
\DeclareFloatingEnvironment[name={Supplementary Figure}]{suppfigure}
\usepackage{sidecap}
\sidecaptionvpos{figure}{c}

\usepackage{titlesec}
\titlespacing\section{0pt}{12pt plus 3pt minus 3pt}{1pt plus 1pt minus 1pt}
\titlespacing\subsection{0pt}{10pt plus 3pt minus 3pt}{1pt plus 1pt minus 1pt}
\titlespacing\subsubsection{0pt}{8pt plus 3pt minus 3pt}{1pt plus 1pt minus 1pt}

\definecolor{lime}{HTML}{A6CE39}
\DeclareRobustCommand{\orcidicon}{
	\begin{tikzpicture}
	\draw[lime, fill=lime] (0,0)
	circle [radius=0.16]
	node[white] {{\fontfamily{qag}\selectfont \tiny ID}};
	\draw[white, fill=white] (-0.0625,0.095)
	circle [radius=0.007];
	\end{tikzpicture}
	\hspace{-2mm}
}
\foreach \x in {A, ..., Z}{\expandafter\xdef\csname orcid\x\endcsname{\noexpand\href{https://orcid.org/\csname orcidauthor\x\endcsname}
			{\noexpand\orcidicon}}
}

\usepackage[ruled,vlined]{algorithm2e}

\usepackage{xcolor}

\usepackage{multirow}
\usepackage{csquotes}
\MakeOuterQuote{"}
\usepackage{hyperref}
\usepackage{xspace}
\newcommand{\mytitle}[1]{\smallskip\noindent\textbf{#1.}\xspace}
\usepackage{graphicx}
\usepackage[toc,page]{appendix}
\usepackage{enumitem}
\setlist{nosep}

\usepackage{subcaption}
\usepackage{multirow}
\usepackage{booktabs}
\usepackage{tabularx}
\usepackage{array}
\usepackage{makecell}
\usepackage{booktabs}
\usepackage{multirow}
\usepackage{siunitx}
\usepackage[T1]{fontenc}
\usepackage[htt]{hyphenat}
\newcommand{\sysname}{dLLM-Serve\xspace}

\title{Taming the Memory Footprint Crisis: System Design for Production Diffusion LLM Serving}

\author{Jiakun Fan, Yanglin Zhang, Xiangchen Li, Dimitrios S. Nikolopoulos}

\begin{document}

\twocolumn[ 
\begin{@twocolumnfalse} 
\maketitle
\begin{abstract}
Diffusion Large Language Models (dLLMs) have emerged as a promising alternative to Autoregressive Models (ARMs), utilizing parallel decoding to overcome sequential bottlenecks. However, existing research focuses primarily on kernel-level optimizations, lacking a holistic serving framework that addresses the unique memory dynamics of diffusion processes in production. We identify a critical "memory footprint crisis" specific to dLLMs, driven by monolithic logit tensors and the severe resource oscillation between compute-bound "Refresh" phases and bandwidth-bound "Reuse" phases. To bridge this gap, we present dLLM-Serve, an efficient dLLM serving system that co-optimizes memory footprint, computational scheduling, and generation quality. dLLM-Serve introduces Logit-Aware Activation Budgeting to decompose transient tensor peaks, a Phase-Multiplexed Scheduler to interleave heterogeneous request phases, and Head-Centric Sparse Attention to decouple logical sparsity from physical storage. We evaluate dLLM-Serve on diverse workloads (LiveBench, Burst, OSC) and GPUs (RTX 4090, L40S). Relative to the state-of-the-art baseline, dLLM-Serve improves throughput by \textbf{1.61$\times$—1.81$\times$} on the consumer-grade RTX 4090 and \textbf{1.60$\times$—1.74$\times$} on the server-grade NVIDIA L40S, while reducing tail latency by nearly \textbf{4$\times$} under heavy contention. dLLM-Serve establishes the first blueprint for scalable dLLM inference, converting theoretical algorithmic sparsity into tangible wall-clock acceleration across heterogeneous hardware. The code is available at https://github.com/chosen-ox/dLLM-Serve.
\end{abstract}
\vspace{0.35cm}
\end{@twocolumnfalse}]



\section{Introduction}
While Autoregressive (AR) LLMs are fundamentally \textit{memory-bandwidth bound} due to the serial loading of KV states for single-token generation, dLLMs introduce a distinct \textit{memory-capacity} and \textit{compute-intensity} bottleneck. By iteratively refining entire sequences via bidirectional attention, dLLMs expose massive intra-sequence parallelism that can saturate GPU compute units. However, this comes at a steep systems cost: the computation of intermediate activations and sequence-wide logits across all tokens creates massive, oscillating activation footprints that shatter the memory assumptions of existing serving engines. Consequently, standard schedulers that optimize for monotonic AR growth fail to manage the dynamic expansion and contraction of dLLM memory demands, leading to severe underutilization or Out-Of-Memory (OOM) crashes.
Recent results even demonstrate faster-than-AR inference when dLLM decoding is restructured for block-wise parallelism and cache reuse (e.g., D2F)~\cite{wang2025diffusionllmsfasterthanarinference}. Beyond speed, controlled studies indicate that masked/absorbing diffusion can be more data-efficient than size-matched AR models under constrained-data regimes, strengthening the case for dLLM deployment when data, not compute, is the bottleneck~\cite{prabhudesai2025diffusionbeatsautoregressivedataconstrained}.

Modern AR LLM inference frameworks combine batching, KV caching, and Algorithm-Hardware Co-design to maximize the throughput.
However, dLLM inference introduces unique constraints: feature drift and parallel decoding that require specialized handling beyond standard AR designs. First, regarding feature drift, the evolving nature of token representations invalidates append-only caching. To address this, recent works have developed robust validity protocols~\cite{ma2025dkvcachecachediffusionlanguage, liu2025dllmcacheacceleratingdiffusionlarge} that selectively refresh or delay cache updates to ensure state consistency without full recomputation.
Second, regarding parallel decoding, simultaneously committing multiple tokens risks violating local dependencies. To resolve this, confidence-aware policies~\cite{wu2025fastdllmv2efficientblockdiffusion, wang2025diffusionllmsfasterthanarinference} have been proposed to dynamically balance intra-sequence parallelism against generation quality.
Together, these contributions have successfully established the algorithmic foundations for correct and efficient dLLM inference.

While the ecosystem for AR LLM serving is robust, featuring mature frameworks~\cite{SLED1, vllm,sglang,sheng2023flexgen} like vLLM that optimize throughput and latency, today’s dLLM stacks still lack a comparable end-to-end serving framework. Existing work such as dLLM-Cache~\cite{liu2025dllmcacheacceleratingdiffusionlarge}, Fast-dLLM~\cite{wu2025fastdllmtrainingfreeaccelerationdiffusion, wu2025fastdllmv2efficientblockdiffusion}, and Sparse-dLLM~\cite{song2025sparsedllmacceleratingdiffusionllms} optimizes isolated components: cache policies, decoding rules, or sparsity patterns. However, no system jointly budgets activation and KV memory under oscillating demand, nor integrates head-level sparsity with a serving scheduler. As a result, existing frameworks either run dLLMs in toy, single-batch settings or encounter OOM and under-utilization when attempting to serve many concurrent requests on production GPUs.


In this work, we explicitly target online dLLM serving for interactive workloads, where a GPU server must handle tens to hundreds of concurrent prompts with low tail latency. However, achieving this scale reveals three unique system characteristics that distinguish with standard AR LLM serving paradigms. First, we identify a "Memory Footprint Crisis" unique to dLLMs: the monolithic logit computation tensor dominates GPU memory, forcing standard systems to reserve massive activation buffers that strictly limit concurrency. Second, the dLLM denoising loop exhibits distinct resource oscillation, alternating between a compute-bound "Refresh" phase which updates full-sequence QKV and a bandwidth-bound "Reuse" phase which only updates activate selected or in-block tokens and uses KV states of remaining tokens. This fluctuating pattern causes severe under-utilization in static schedulers. Third, standard sparse attention methods are inefficient for serving; they rely on logical masking that wastes physical memory and enforce uniform masks across all heads to simplify implementation, sacrificing model accuracy.

To address these challenges, we present \sysname, a holistic serving system designed for the unique physics of dLLMs. \sysname resolves the massive memory footprint of computation logit tensor via Logit Decomposition, a runtime technique that splits the monolithic output projection into small serial sub-batches along the token axis, strictly bounding the peak activation footprint. To exploit workload heterogeneity, we introduce a Phase-Multiplexed Scheduler that tessellates execution phases, dynamically injecting new requests into the resource headroom released by active requests which transit from refresh phase to reuse phase to maximize hardware utilization. Finally, \sysname implements a Head-Centric Sparse KV Cache with a dense storage layout. This decouples logical sparsity from physical placement, enabling the system to support accurate, irregular, per-head token retention while maintaining fully coalesced memory accesses. To our knowledge, \sysname is the first serving framework to explicitly address the oscillating memory profile of iterative denoising.


Our key contributions are summarized as follows:
\begin{itemize}
  \item We present \sysname, an efficient dLLM serving system that elevates fine-grained memory management to a core design goal.
  \item We propose Logit-Aware Activation Budgeting that decomposes the monolithic logit tensor into serial sub-batches to strictly cap peak activation demands, effectively converting transient memory overhead into persistent KV capacity for higher serving concurrency.
  
  \item We design a Phase-Multiplexed scheduler that orchestrates the execution of compute-bound 'Refresh' and bandwidth-bound 'Reuse' phases to maximize global hardware utilization.

  \item We implement a Head-Centric KV Cache Memory Management that decouples logical sparsity from physical placement to support irregular, per-head token retention. This design preserves high model accuracy while packing data into dense buffers to ensure fully coalesced memory accesses for maximum bandwidth efficiency.

  \item We evaluate our \sysname with state-of-the-art dLLM baselines. Our evaluation demonstrates that our techniques achieve up to $1.81\times$ throughput improvement over state-of-the-art dLLM baselines while maintaining high generation quality.

\end{itemize}
\section{Background}
\label{sec:background_motivation}

\subsection{Autoregressive LLMs}


Autoregressive language models generate text by factorizing the joint distribution into a product of conditional distributions:
\begin{equation}
p_\theta(\mathbf{x}_{1:L}) = \prod_{t=1}^{L} p_\theta(x_t \mid \mathbf{x}_{<t}),
\end{equation}
where $\mathbf{x}_{1:L}$ denotes a sequence of length $L$, and $x_t$ is the token at position $t$. During inference, tokens are generated sequentially from left to right: $x_t \sim p_\theta(\cdot \mid \mathbf{x}_{1:t-1})$ for $t=1,\ldots,L$. This requires $L$ sequential forward passes through the model, where each step conditions only on previously generated tokens via a causal attention mask.  

AR LLM inference proceeds in two phases: prefill, where the model processes the initial prompt and stores the KV cache to prevent redundant recomputation, and decode, where the model generates tokens sequentially by attending to the prefix KV cache. As the sequence grows, the KV cache footprint expands linearly, making memory capacity the primary bottleneck for serving throughput. To address this, modern serving engines employ Continuous Batching and PagedAttention~\cite{vllm} to minimize fragmentation and maximize memory utilization. Furthermore, the FlashAttention series~\cite{dao2022flashattention, dao2023flashattention2} optimizes the attention kernel itself, enabling fast, memory-efficient exact attention through IO-aware tiling.

 While AR model is simple and effective, the sequential dependency prevents parallelization across the sequence length dimension, and the unidirectional context limits the model's ability to leverage future information during generation.

\subsection{Diffusion LLMs Inference}
In contrast to AR LLMs, diffusion large language models (dLLMs) like LLaDA generate text through an iterative denoising process that predicts all tokens in parallel. Starting from a fully masked sequence $\mathbf{x}_1 = [\texttt{MASK}]^L$, the model performs $K$ denoising steps, where at each step from time $\tau$ to $\tau - \delta$:
\begin{equation}
x_{\tau-\delta}^i \sim p_\theta(x_0^i \mid \mathbf{x}_\tau) \quad \text{for all } i \text{ where } x_\tau^i = \texttt{MASK},
\end{equation}
followed by remasking a fraction of the predicted tokens. Here, $p_\theta(x_0^i \mid \mathbf{x}_\tau)$ denotes the probability of the clean token at position $i$ given the current masked sequence $\mathbf{x}_\tau$, and the model uses bidirectional attention to condition on the entire context. This allows all masked positions to be decoded simultaneously in each step. The bidirectional modeling enables the network to incorporate both past and future context, which can improve performance on tasks requiring non-causal reasoning, though it trades the sequential dependency of autoregressive models for an iterative refinement process.

However, dLLM inference challenges the established AR LLM inference designs due to the unique constraints of bidirectional attention and iterative denoising.
First, the evolving nature of token representations invalidates standard caching assumptions. Unlike autoregressive models where past tokens remain fixed, dLLM tokens are updated at every denoising step. Consequently, "naive" append-only caching which assumes static history leads to \textit{feature drift}, where stored KV states diverge from the current denoising trajectory, accumulating significant error. To mitigate this, recent systems employ strictly controlled reuse protocols, such as one-step delayed reuse in dKV-Cache~\cite{ma2025dkvcachecachediffusionlanguage} or adaptive, feature-aware refresh policies in dLLM-Cache~\cite{liu2025dllmcacheacceleratingdiffusionlarge} that selectively recompute volatile states.
Second, the ability to commit multiple tokens per step introduces a complex trade-off between parallelism and correctness. While dLLMs can predict parallel blocks, aggressive multi-token commitment can violate local dependencies and degrade coherence. Optimizing this requires confidence-aware policies (e.g., Fast-dLLM~\cite{wu2025fastdllmv2efficientblockdiffusion}, D2F~\cite{wang2025diffusionllmsfasterthanarinference}) that dynamically balance intra-sequence parallelism against generation quality.
Finally, the prevalence of mask tokens with low information density renders dLLMs highly amenable to sparsity, motivating dynamic management strategies like attention-guided eviction (Sparse-dLLM~\cite{song2025sparsedllmacceleratingdiffusionllms}) or head-specific sparse patterns (SparseD~\cite{wang2025sparsedsparseattentiondiffusion}) to concentrate limited GPU capacity on high-importance tokens. Together, these algorithmic adaptations shape unique memory footprints, presenting specific patterns that can be exploited to optimize dLLM serving efficiency.

\subsection{Two-Phase dLLM Inference Process}
\label{sec:background:execution_model}

To formally characterize the workload characteristics of dLLMs, we model the inference process as a sequence of discrete timesteps $\mathcal{T}=\{T,T-1,\dots,1\}$. For block-wise (semi-autoregressive) generation, tokens are decoded in blocks of size $B_{\text{size}}$ at each time step. 
To mitigate cache staleness and prevent error propagation, most dLLM algorithms introduced earlier adopt an iterative inference process comprising Refresh and Reuse phases, thereby balancing efficiency with accuracy.

\mytitle{Refresh Phase}
When transitioning between consecutive blocks or blocks with a fixed interval $K_{\text{int}}$, the dLLM synchronizes and updates the KV cache for the whole sequence of $L$ tokens with the current hidden states in this step, which is called Refresh phase. Specifically, in a Refresh phase, dLLM computes attention and update QKV ($Q_f,K_f,V_f$) for the full sequence according to the role shown below:

\begin{equation}
O_f
=
\text{Softmax}!\left(
\frac{Q_f K_f^\top}{\sqrt{d}}
\right)
V_f.
\end{equation}

\mytitle{Reuse Phase} 
Within a block window, the context outside the active block is treated as static. The system reuses the cached context keys/values $(K_{\text{cache}}, V_{\text{cache}})$ of tokens outside the active block and computes attention only for the active block in this step, which is called Reuse phase. The computation of attention is shown below:
\begin{equation}
O_b
=
\text{Softmax}!\left(
\frac{Q_b [K_b; K_{\text{cache}}]^\top}{\sqrt{d}}
\right)
[V_b; V_{\text{cache}}].
\end{equation}

This two-phase inference induces temporal variations in the memory footprint across iterative steps. Such fluctuations should be leveraged in serving system to optimize concurrency and maximize throughput.

\subsection{Sparse Attention for dLLMs}
\label{sec:background:sparsity}

Diffusion LLMs operate via iterative denoising, where a substantial portion of the sequence is comprised of mask tokens. Because these tokens carry negligible semantic information, the attention distribution is inherently sparse, focusing primarily on the unmasked, informative regions. This characteristic makes dLLMs particularly well-suited for sparse attention mechanisms. By filtering out low-information mask tokens and addressing the quadratic $\mathcal{O}(N^2)$ complexity, sparse algorithms mitigate compute and memory traffic by restricting attention computation to a subset of "salient" context tokens.

Current methods such as Sparse-dLLM~\cite{song2025sparsedllmacceleratingdiffusionllms} enforce a single global sparsity pattern to preserve memory contiguity and kernel regularity. Concretely, the system ranks context tokens using an importance score that is aggregated across heads. To jointly capture (i) local neighborhood structure, where relevance may shift within a short window, and (ii) global heavy-hitters that remain consistently useful, these methods apply a local pooling operator with kernel size $w$ to the raw per-head dot-product scores. The resulting sequence-level importance score for context position $j$ is:
\begin{equation}
S_j = \sum_{h=1}^H \left( \max_{m \in [j-\frac{w}{2},, j+\frac{w}{2}]} (Q_{b,h} \cdot K_{m,h}^\top) \right).
\end{equation}
Here, the inner $\max(\cdot)$ performs local pooling: a token receives credit if any nearby position in a window around $j$ is strongly aligned with the block query under head $h$. The outer summation then aggregates evidence across all $H$ heads, producing a single global score vector $S \in \mathbb{R}^N$.

The system selects one shared top-$k$ index set $\mathcal{I}$ from $S$, and uses $\mathcal{I}$ to gather the retained keys/values for \emph{every} head. Because all heads share $\mathcal{I}$, the KV gathers exhibit a regular, head-aligned access pattern: the retained KV slices can be laid out (or fetched) in a naturally contiguous manner, enabling coalesced loads and avoiding highly irregular per-head indirections in standard attention kernels.

However, the same uniformity is also the central limitation. By forcing all heads to attend to the same retained context, sequence-level sparsity implicitly optimizes for \emph{average} relevance across heads, which can discard tokens that are crucial for specialized semantic heads but appear weak when aggregated. This head-agnostic selection can therefore trade accuracy for efficiency, especially at aggressive retention ratios. While head-level sparsity, where token selection varies across heads, resolves this by preserving distinct semantic dependencies, it requires efficient KV cache management, enabling the system to support per-head selection while maintaining high memory efficiency during serving.

\section{Motivation}

\newcolumntype{Y}{>{\raggedright\arraybackslash}X}
\begin{table*}[t]
\centering
\caption{\textbf{Qualitative Comparison with State-of-the-Art.} Unlike prior works that focus on isolated caching policies or attention sparsity algorithms, \sysname is a holistic serving system. It introduces fine-grained \textit{phase-level scheduling} and \textit{continuous batching} to the dLLM domain. In the footprint column, $L$ denotes the total request sequence length and $r$ represents the sparsity retention ratio ($r < 1$).}
\label{tab:comparison}
\scriptsize
\setlength{\tabcolsep}{6pt}
\renewcommand{\arraystretch}{1.12}

\begin{tabularx}{\linewidth}{@{} l c c c Y Y @{}}
\toprule
\textbf{Work} &
\textbf{System Scope} &
\textbf{Batching Strategy} &
\makecell[c]{\textbf{Scheduling Granularity}} &
\textbf{KV Cache Scope} &
\makecell[l]{\textbf{Cache Footprint Per Layer}} \\
\midrule
dKV-Cache~\cite{ma2025dkvcachecachediffusionlanguage} &
Cache Policy & Static & Request-Level &
Decoded tokens (delayed) &
$\approx L \times$ sizeof(Key + Value) \\

dLLM-Cache~\cite{liu2025dllmcacheacceleratingdiffusionlarge} &
Cache Policy & Static & Request-Level &
Prompt + Adaptive Response &
$\approx L \times$ sizeof(KV + Attn + FFN) \\

Fast-dLLM~\cite{wu2025fastdllmv2efficientblockdiffusion} &
Inference Framework & Static & Request-Level &
Prefix / Suffix Blocks &
$\approx L \times$ sizeof(Key + Value) \\

Sparse-dLLM~\cite{song2025sparsedllmacceleratingdiffusionllms} &
Sparsity Algo & Static & Request-Level &
Global Top-$k$ Tokens (Shared) &
$\approx rL \times$ sizeof(Key + Value) \\

SparseD~\cite{wang2025sparsedsparseattentiondiffusion} &
Sparsity Algo & Static & Request-Level &
No Cache; Logical Masks Only &
None (Stateless) \\

dInfer~\cite{ma2025dinferefficientinferenceframework} &
Inference Framework & Dynamic & Request-Level &
Vicinity Refresh &
$\approx L \times$ (Key + Value) \\
\midrule
\textbf{\sysname (Ours)} &
\textbf{Full Serving System} &
\textbf{Continuous (Paged)} &
\textbf{\makecell[c]{Phase-Level (Refresh/Reuse)}} &
\textbf{\makecell[l]{Per-Head Top-$k$ Tokens}} &
\textbf{\makecell[l]{$\approx rL \times$ sizeof(Key + Value);\\Stored contiguously}} \\
\bottomrule
\end{tabularx}
\end{table*}

While recent algorithmic works optimize single-request latency, they fail to address the serving-layer constraints required for high-throughput concurrency.

\subsection{Static Request-Level Scheduling Wastes Phase Heterogeneity}
\label{sec:motivation:scheduling}

The most distinct system characteristic of modern dLLM inference is the violent resource oscillation between denoising steps. As described in \S\ref{sec:background:execution_model}, the workload alternates between two extremes: Refresh phase and Reuse phase.
The Refresh phase dominates peak memory and calculating global attention. In contrast, the Reuse phase is lightweight in compute but bandwidth-bound due to loading cached KV states.

\mytitle{The Granularity Mismatch}
Despite this predictable fluctuation, existing frameworks~\cite{ma2025dinferefficientinferenceframework, wu2025fastdllmv2efficientblockdiffusion} typically schedule at the coarse granularity of a request. They treat the internal phase transitions as a black box and must conservatively provision resources for the Refresh phase throughout the request's entire lifetime.
Consequently, during the bandwidth-bound Reuse phases where activation and compute demands drop precipitously, the massive reserved budget sits idle. These systems cannot systematically exploit this "Reuse headroom" to admit additional work, resulting in low average GPU utilization and suboptimal throughput. \sysname addresses this with a Phase-Multiplexed Greedy Scheduler (\S\ref{sec:design:scheduler}) that schedules at the granularity of individual phases.

\subsection{Unmanaged Peak Activations Constrain Concurrency}
\label{sec:motivation:activation}

The second major bottleneck lies in the massive, transient memory footprint of the output projection. In standard AR decoding, the final layer produces logits only for the last token position, yielding a small tensor of shape $[B,1,V]$. In contrast, dLLMs naturally produce logits for the entire active region of length $L$ to refine the sequence iteratively, yielding a monolithic tensor of shape $[B,L,V]$.

\mytitle{The Logit-Memory Boom}
This difference shifts the memory bottleneck from the  persistent KV cache to transient activations. For a representative LLaDA-8B configuration ($B{=}16$, $L{=}2048$, $V{=}126{,}464$, FP16), the logit tensor alone requires:
\[
16 \times 2048 \times 126{,}464 \times 2\ \text{bytes} \approx 8.3\ \text{GB}.
\]
This single tensor is transient, living only for the duration of the output projection, yet it determines the peak memory requirement. If the serving system does not account for this spike, a single batched step can trigger an Out-Of-Memory (OOM) error.

\mytitle{Lack of Explicit Budgeting}
Current optimizations do not address this "activation boom." Methods like Fast-dLLM and dKV-Cache~\cite{wu2025fastdllmv2efficientblockdiffusion, ma2025dkvcachecachediffusionlanguage} focus on reducing FLOPs but still materialize the full $[B,L,V]$ tensor when they \textit{do} execute a step.Similarly, frameworks like dInfer~\cite{ma2025dinferefficientinferenceframework} optimize execution graphs but leave the logit tensor exposed.
Without an explicit mechanism to decompose this tensor, the serving engine must act conservatively which limits the batch size $B$ to accommodate the worst-case allocation of the monolithic projection. This artificial cap on concurrency leaves GPU compute underutilized during the rest of the forward pass. 
\sysname explicitly manages this via Logit-Aware Activation Budgeting (\S\ref{sec:design:budgeting}).

\subsection{Logical Sparsity Fails to Reclaim Physical Resources}
\label{sec:motivation:sparsity}

To further accelerate dLLM inference, recent research has explored sparse attention for dLLMs~\cite{song2025sparsedllmacceleratingdiffusionllms}. However, translating theoretical sparsity into system-level gains remains an open challenge due to a disconnect between logical patterns and physical storage.

\mytitle{The Uniformity Trap}
Current sparse methods such as Sparse-dLLM~\cite{song2025sparsedllmacceleratingdiffusionllms} enforce a \emph{uniform mask} across all attention heads to simplify memory management. While this preserves memory contiguity, it implicitly optimizes for \textit{average} relevance. This is detrimental for dLLMs, where different heads often specialize in distinct semantic features. Forcing all heads to attend to the same subset of tokens discards context that is crucial for specific heads, degrading generation quality at aggressive retention ratios.

\mytitle{The Physical Layout Gap}
While recent works like SparseD~\cite{wang2025sparsedsparseattentiondiffusion} demonstrate that \emph{head-specific} patterns can restore this quality, they lack a serving-grade implementation. They typically realize sparsity via logical masking.
This approach does not save physical memory capacity or reduce bandwidth usage; the system still allocates the full KV buffer and loads irrelevant tokens, only to mask them out during computation. Without a physical data layout that can store irregular, per-head selections contiguously, the theoretical efficiency of sparse attention cannot be realized in a high-throughput serving engine.  \sysname closes this gap with Head-Centric Sparse KV Cache Management (\S\ref{sec:design:sparsity}).

Table~\ref{tab:comparison} summarizes these gaps. \sysname is the only end-to-end serving system that combines continuous batching with phase-level scheduling to handle oscillating activation memory demands, and a coalesced head-centric KV organization with minimized footprint designed for multi-request serving.

\begin{figure*}[t]
    \centering
    \includegraphics[width=\textwidth]{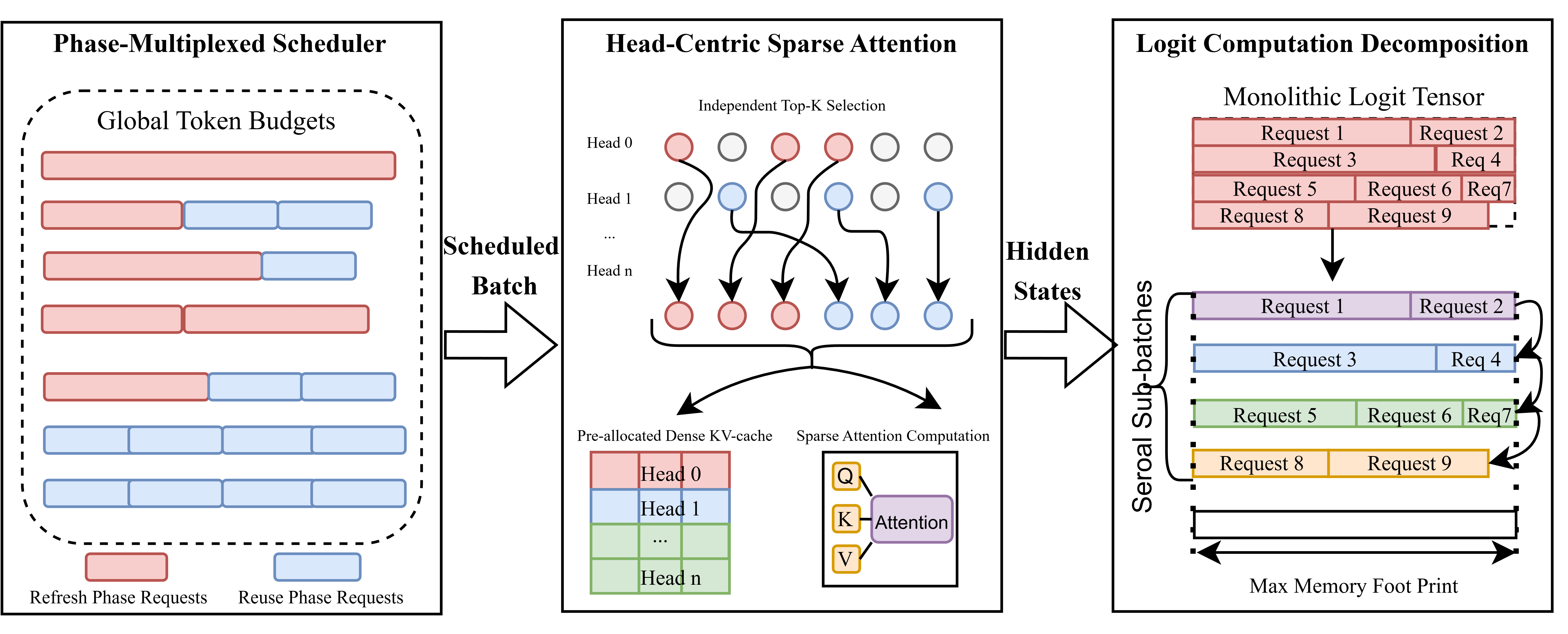}
    \caption{\sysname Overview of \sysname.}
    \label{fig:arch}
\end{figure*}

\section{System Design}
\label{sec:design}

We present \sysname, a serving framework co-designed with the unique memory properties of diffusion models. Unlike autoregressive systems that optimize for strictly monotonic sequence growth, \sysname manages the oscillating memory demands of iterative denoising.


\subsection{System Overview}
\label{sec:design:overview}

Figure~\ref{fig:arch} illustrates the high-level workflow of \sysname, which transforms the irregular resource patterns of dLLM inference into a structured, schedulable pipeline. The system operates through four coordinated stages:

\begin{enumerate}[leftmargin=*]

\item \textbf{Initialization \& Budgeting.}
On startup, the Offline Profiler (\S\ref{sec:design:profiler}) measures a conservative upper bound on the activation footprint under the user configuration.
Crucially, profiling incorporates Logit Decomposition (\S\ref{sec:design:budgeting}) by enforcing \texttt{max\_num\_logits} during the profile run. This reserves a smaller activation region and increases the remaining HBM available for the KV Cache Pool.


\item \textbf{Phase-Aware Scheduling.}
Incoming requests enter a queue managed by the Phase-Multiplexed Greedy Scheduler (\S\ref{sec:design:scheduler}). Each iteration, the scheduler constructs a single packed token batch whose size is bounded by \texttt{max\_num\_batched\_tokens}. Requests currently in Refresh contribute many query tokens, while requests currently in Reuse contribute only a small block of query tokens. This allows \sysname to admit new Refresh work exactly when other requests enter Reuse and release token/activation headroom.


\item \textbf{Sparse KV Management.}
Once scheduled, the Head-Centric Memory Manager (\S\ref{sec:design:sparsity}) allocates KV pages from the global KV pool and exposes them as a logically dense packed buffer for attention. The cache footprint is fixed per request, enabling predictable admission control while preserving head-specific selection.


\item \textbf{Execution with Logit Decomposition.}
The packed batch executes on the GPU. If the number of tokens requiring logits in this iteration exceeds \texttt{max\_num\_logits}, \sysname computes logits in serial sub-batches (\S\ref{sec:design:budgeting}) so that the runtime peak activation footprint never exceeds the profiled activation budget.

\end{enumerate}

The following subsections detail these core mechanisms. We begin with Logit-Aware Budgeting (\S\ref{sec:design:budgeting}), which establishes the memory foundation for concurrency. We then describe the Phase-Multiplexed Scheduler (\S\ref{sec:design:scheduler}) that exploits this capacity, and finally the Head-Centric Memory Manager (\S\ref{sec:design:sparsity}) that ensures high-quality generation under strict memory constraints.

\begin{figure}[h!]
    \centering
\includegraphics[width=0.8\columnwidth]{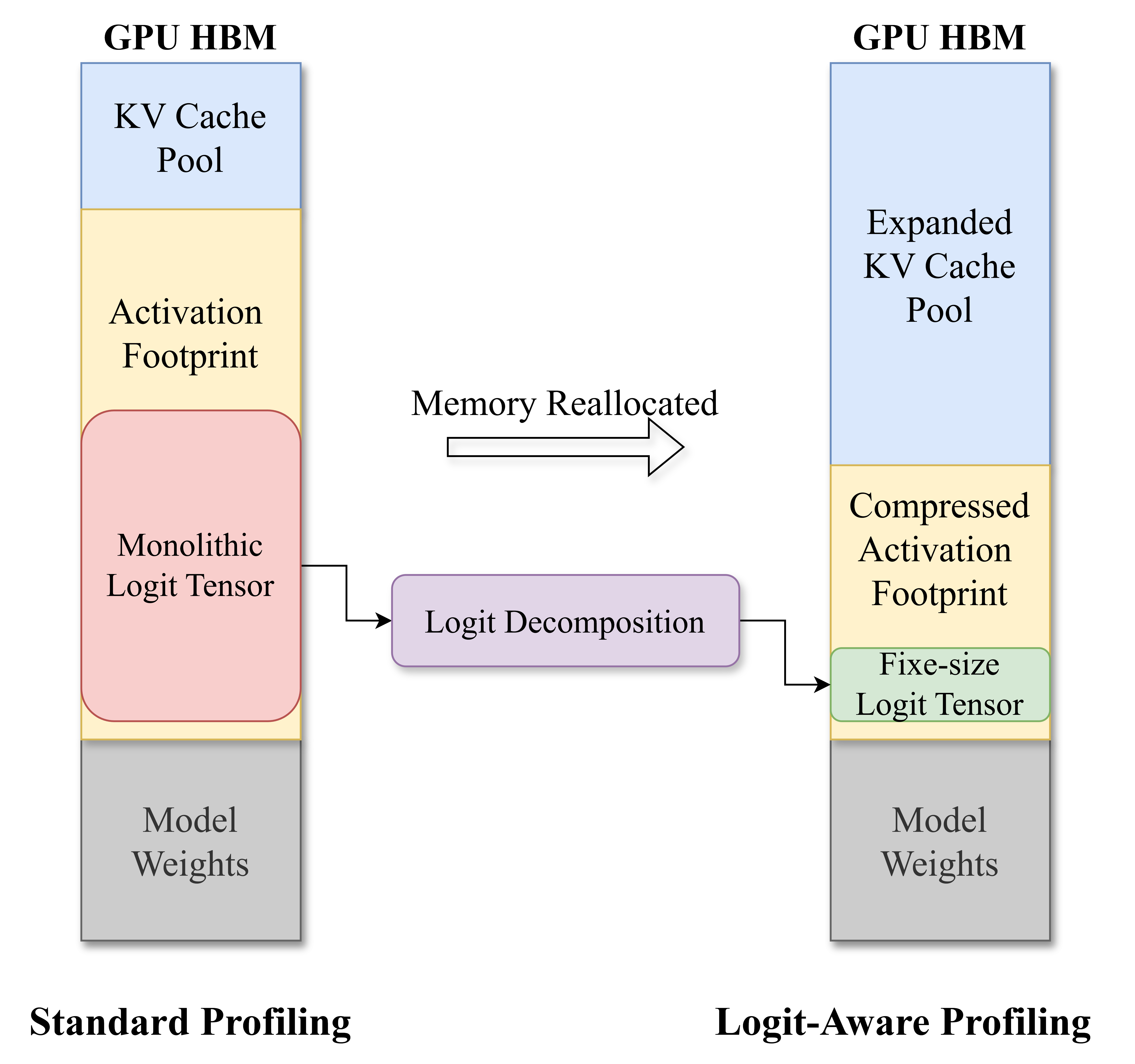}
    \caption{Standard profiling vs. logit-aware profiling.}
    \label{fig:single}
\end{figure}
\subsection{Offline Memory Profiler}
\label{sec:design:profiler}
\sysname begins with an initialization phase that maps the GPU memory envelope under worst-case serving pressure. Following the methodology used in vLLM~\cite{vllm}, the profiler executes dummy workloads under the configured \texttt{max\_num\_batched\_tokens} to measure a conservative peak activation footprint for intermediate tensors (e.g., attention/MLP workspaces and GEMM buffers). We add a small guard band to account for allocator fragmentation and kernel selection variability.

In contrast to Auto-Regressive (AR) model profiling, dLLM serving necessitates explicit accounting for output-projection activations. Consequently, \sysname integrates \texttt{max\_num\_logits} constraints (via Logit Decomposition; \S\ref{sec:design:budgeting}) directly into its profiling logic. As illustrated in Figure~\ref{fig:single}, standard profiling strategies must reserve substantial activation memory to accommodate monolithic logit computation, thereby severely limiting the capacity of the KV pool. Conversely, \sysname's logit-aware profiling exploits logit decomposition to minimize activation reservation, effectively liberating memory to expand the KV pool and enhance system concurrency.

\subsection{Logit-Aware Activation Budgeting}
\label{sec:design:budgeting}

A critical scalability bottleneck in dLLM serving is the massive activation footprint of the output projection. While efficient decoding algorithms may only require updating tokens within a specific block, existing serving systems often lack the granularity to exploit this. Naive implementations typically run the Transformer in a monolithic mode, materializing logits for the full sequence length ($L$) at every step.
This results in a massive tensor $\mathbf{Z} \in \mathbb{R}^{B \times L \times V}$.
For large batch sizes, this memory explosion dominates HBM usage, forcing the scheduler to drastically limit concurrency to avoid Out-Of-Memory (OOM) errors during peak allocation.

\mytitle{Logit Decomposition}
To break this dependency, \sysname implements runtime token-axis splitting.
We introduce a system parameter \texttt{max\_num\_logits} that defines the maximum number of tokens for which logits can be materialized simultaneously.
During execution, let $N_{\text{logit}}$ be the total number of tokens requiring logit computation in the current batch.
If $N_{\text{logit}} > \texttt{max\_num\_logits}$, the runtime decomposes the output projection into serial sub-batches.

The engine iterates through these chunks, processing at most \texttt{max\_num\_logits} tokens at a time. For each chunk, it computes the logits, immediately applies the decoding operator (e.g., ArgMax or Top-$k$ sampling) to obtain the next tokens, and crucially, releases the temporary logit buffer before proceeding to the next chunk.
This ensures that the instantaneous activation footprint is strictly bounded by the chunk size, regardless of the full sequence length or batch size.

\mytitle{KV Cache Maximization}
This decomposition directly translates to higher serving capacity. Because the Offline Profiler (\S\ref{sec:design:profiler}) is aware of the \texttt{max\_num\_logits} constraint, it reserves a strictly minimized activation budget.
The gigabytes of HBM reclaimed from preventing the naive "activation spike" are reallocated to the KV Cache Pool.
Consequently, \sysname can accommodate a significantly larger number of concurrent requests and their associated KV cache within the same fixed memory budget, converting transient activation waste into persistent serving throughput.

\subsection{Phase-Multiplexed Greedy Scheduler}
\label{sec:design:scheduler}

Many existing dLLM runtimes schedule at request granularity and implicitly provision for peak cost throughout the request lifetime. \sysname instead schedules at step granularity using token-level packing, governed by \texttt{max\_num\_batched\_tokens}.

\mytitle{Scheduling Currency and Invariant}
The scheduler enforces a strict invariant: the total number of active \emph{query} tokens in the packed batch never exceeds \texttt{max\_num\_batched\_tokens}. We use query tokens as the scheduling currency because the per-iteration activation workspace for attention/MLP scales with the number of query tokens processed, while (i) KV resides in the pre-allocated KV pool and (ii) logits are bounded separately by \texttt{max\_num\_logits}.

\mytitle{Phase-Aware Batching}
In each denoising iteration, \sysname constructs a single packed batch of active tokens across requests. A request in the Refresh phase contributes query tokens of the full sequence length, while a request in the Reuse phase contributes query tokens for only the active block. The runtime executes one forward pass over this packed batch using FlashAttention’s variable-length interface, with per-request metadata providing the appropriate KV pointers and lengths.

\mytitle{Greedy FCFS Admission}
We implement a greedy FCFS policy. When running requests transition from the Refresh phase to the Reuse phase, their contribution to the packed batch drops from $L_{\text{total}}$ to $L_{\text{block}}$, freeing significant query-token budget. The scheduler immediately admits queued requests starting in the Refresh phase until the packed batch again reaches \texttt{max\_num\_batched\_tokens}. This converts the headroom created by bandwidth-dominated Reuse phases into additional compute-heavy refresh work, increasing overall utilization.

\subsection{Head-Centric Sparse KV Cache Management}
\label{sec:design:sparsity}
To reconcile dLLM memory pressure with generation quality, \sysname implements a Head-Centric Sparse KV cache. The goal is to support head-specific token retention while preserving a contiguous physical layout for GPU-efficient attention.



\mytitle{Algorithmic formulation}
Unlike global sparsity methods (\S\ref{sec:background:sparsity}) that aggregate scores across heads to enforce a shared mask, \sysname computes independent per-head importance scores using local pooling (kernel size $w$):

\begin{equation}
    S_{h,j} =
    \max_{m \in [j-\frac{w}{2},\, j+\frac{w}{2}]}
    (Q_{b,h} \cdot K_{m,h}^\top),
\end{equation}
and selects $\mathcal{I}^h = \text{TopK}(S_{h,:}, k)$ with $k=L\cdot r$. In general $\mathcal{I}^h \neq \mathcal{I}^{h'}$, which improves modeling capacity but induces divergent gather operations under standard KV cache layouts.

\mytitle{Decoupling Logical Sparsity from Physical Layout}
\sysname eliminates the latency of divergent gather operations during the Reuse phase by decoupling token selection from the physical storage layout. During the Refresh phase, the system uses the selected indices $\mathcal{I}^h$ to immediately pack the sparse tokens into a physically dense KV layout. Crucially, the sparse index map is transient; it is used solely for this packing step and is not maintained in the KV cache. This decoupling allows the subsequent Reuse phase to access the KV cache sequentially from contiguous memory, completely bypassing the need for indirect addressing.

\mytitle{Static Allocation and Contiguous Storage}To support this packed execution, the memory manager enforces a strict static budget. For each request with sequence length $L$ and retention ratio $r$, the system pre-allocates a fixed block of $rL \times \text{sizeof}(\text{KV})$ memory, organizing it into a contiguous dense tensor of shape $[N_{\text{heads}}, rL, D_{\text{head}}]$. This strategy guarantees a physically dense storage layout throughout the request lifetime. Furthermore, because positional information is embedded into the keys prior to storage, the Reuse phase executes attention directly over this contiguous buffer without the overhead of position re-calculation.

\section{Implementation}
\label{sec:implementation}


We implemented \sysname by extending the open-source \texttt{Nano-vLLM}~\cite{nano-vllm} runtime. The implementation adds diffusion-specific control flow including iterative denoising, phase tracking, and sparse KV cache management while reusing the high-performance components already present in \texttt{Nano-vLLM}~\cite{nano-vllm}, including paged KV cache storage and FlashAttention. The system is implemented in PyTorch and integrates with HuggingFace Transformers.

\subsection{dLLM Execution Runtime}


\texttt{Nano-vLLM}~\cite{nano-vllm} is designed for AR LLMs with a single forward pass per decoding step. We extend its \texttt{ModelRunner} into a diffusion-aware runtime that executes an iterative denoising process and exposes the phase information required by our scheduler (\S\ref{sec:design:scheduler}).

\mytitle{Iterative Denoising Loop}
We replace the AR-style step function with a stateful denoising loop. At each denoising step $t$, the runner injects the diffusion time-step embedding, executes the Transformer backbone, applies the mask-prediction logic, and updates the input token tensor for the next step. To reduce Python overhead and allocator churn, we (i) precompute the diffusion noise schedule and time-step embeddings on GPU during initialization, and (ii) update token IDs in-place when transitioning masked positions to generated tokens.


\mytitle{Logit Decomposition (token-axis microbatching)}
To enforce the budgeting policy (\S\ref{sec:design:budgeting}), we refactor the LM head into a sliced execution path controlled by \texttt{max\_num\_logits}. It packs the corresponding hidden states across the batch and splits the LM head in micro-batches. Each micro-batch computes logits, derives the decoding decision, and releases the temporary logits and normalization buffers before processing the next micro-batch. This bounds peak allocation from the output stage while preserving high concurrency.



\subsection{Phase-Aware Control Plane}
\label{sec:implementation:control}

We extend \texttt{nano-vllm}'s scheduler to support dLLM requests lifecycles. Unlike AR serving (prefill then decode), dLLM requests iterate over denoising steps and alternate between Refresh and Reuse phases.

\mytitle{State tracking}
We augment per-request metadata with the current denoising step, the block position, and a phase flag derived from the cache policy such as whether the request should refresh or reuse cache at the current step. The control plane maintains a waiting queue for new requests and a running set for active denoising requests.

\mytitle{Dynamic cost model}
We implement the phase-dependent token accounting from \S\ref{sec:design:scheduler}. In each scheduling iteration, the scheduler computes each running request's active-query contribution: full-sequence in Refresh phase, block-size in Reuse [hase, and admits additional waiting requests whenever running requests transition into Reuse and release budget under \texttt{max\_num\_batched\_tokens}.

\subsection{Sparse Attention Integration}
\label{sec:implementation:sparse}

\sysname supports head-centric sparsity while retaining FlashAttention as the attention kernel. Requests within the same packed batch may be heterogeneous: some execute Refresh phase with dense context, while others execute Reuse phase over a packed sparse history.

\mytitle{Input adaptation for FlashAttention}
We implement a lightweight assembly layer that constructs a FlashAttention-compatible view of each request's effective Q/K/V. During Refresh phase, the runtime computes per-head token importance and packs the retained KV entries into a physically contiguous layout backed by paged KV storage. During Reuse phase, the runtime loads the packed sparse history and combines it with the current block state to form the effective attention input. The assembled per-request segments are then flattened into variable-length Q/K/V buffers with associated offset metadata, enabling a single variable-length FlashAttention dispatch for the entire batch.

\section{Evaluation}
\subsection{Experiment Setup}
\mytitle{Testbed} We run our experiments on two platform: dual Intel Xeon(R) Platinum 8375C (2 * 32 cores) with an NVIDIA GeForce RTX 4090 and Intel Xeon(R) Platinum 8462Y+ (32 cores) with an NVIDIA L40s-48G PCIe. The hardware characteristics of our experiment settings are shown in Table~\ref{tab:hw}.

\begin{table}
  \caption{Hardware Settings of Our Experiments}
  \begin{tabular}{ccl}
    \toprule
    CPU(cores) & RAM & GPU \\
    \midrule
     2 Xeon(R) Platinum 8375C & 250GB & NVIDIA RTX 4090  \\
     Xeon(R) Platinum 8462Y+ & 500GB & NVIDIA L40s-48G PCIe  \\
  \bottomrule
\end{tabular}
\label{tab:hw}
\end{table}

\mytitle{Models \& Memory Constraints.} We evaluate the LLaDA-8B-Instruct model across both platforms to cover distinct operating regimes. We selected this specific model-hardware pairing because the LLaDA-8B architecture requires approximately 17GB of VRAM just to store weights in half-precision (FP16), rendering standard 16GB enterprise cards (e.g., Tesla T4) insufficient for serving. Consequently, our dual-platform setup captures two critical scenarios: the RTX 4090 (24GB) represents a memory-constrained environment where, after loading weights, less than 7GB remains for the KV cache and activations, forcing the system to rely heavily on our efficient budgeting and sparsity mechanisms. In contrast, the NVIDIA L40S (48GB) represents a high-capacity server environment with nearly 31GB of available headroom, allowing us to evaluate the system's scalability and batching efficiency when memory capacity is not the immediate bottleneck.

\mytitle{Baseline}
We compare \sysname with 3 baselines listed below:
\begin{itemize}[leftmargin=*]
    \item \textbf{Fast-dLLM} is a diffusion-based Large Language Model (LLM) inference acceleration framework that supports efficient inference for models like Dream and LLaDA.
    \item \textbf{dLLM-Cache} is a training-free adaptive caching framework that combines long-interval prompt caching with partial response updates guided by feature similarity.
    \item \textbf{Sparse-dLLM} is the first training-free framework integrating dynamic cache eviction with sparse attention via delayed bidirectional sparse caching.
\end{itemize}

\begin{table}[t]
\centering
\caption{\textbf{System Hyperparameters \& Configuration.}}
\label{tab:system_config}
\resizebox{\columnwidth}{!}{%
\begin{tabular}{l|l|c}
\toprule
\textbf{Category} & \textbf{Parameter} & \textbf{Value} \\
\midrule
\multirow{3}{*}{\textbf{Global Generation}} 
 & Generation Length & 256 tokens \\
 & Total Decoding Steps & 256 \\
 & KV Block Size & 32 \\
\midrule
\multirow{2}{*}{\textbf{Sparsity Model}}
 & Retention Ratio & 0.5 \\
 & Kernel Size & 3 \\
\midrule
\multirow{3}{*}{\textbf{Baselines}}
 & Fast-dLLM Mode & Dual-Cache (No Parallel Dec.) \\
 & dLLM-Cache Prompt Interval & 100 steps \\
 & dLLM-Cache Gen Interval & 7 steps \\
\midrule
\multirow{3}{*}{\textbf{\sysname (Ours)}}
 & Max Batched Tokens (4090) & 4,000 \\
 & Max Batched Tokens (L40S) & 16,384 \\
 & Max Materialized Logits & 2,048 \\
\bottomrule
\end{tabular}
}
\end{table}

\mytitle{System Configuration}
 We standardize generation parameters across all systems to ensure fairness. Table~\ref{tab:system_config} summarizes the hyperparameters used across all experiments. To strictly isolate the performance impact of our system-level contributions, we enforce three critical fairness constraints:

\begin{itemize}[leftmargin=*]
    \item \textbf{Algorithmic Consistency:} We fix the retention ratio at $0.5$ which is the default value of \textit{Sparse-dLLM} for both \sysname and \textit{Sparse-dLLM}. While \sysname's head-level selection can maintain accuracy at lower sparsity levels, we align these parameters to ensure that any observed speedups stem from our serving engine and scheduler, not from aggressive retention ratio.
    \item \textbf{Execution Parity:} For the \textit{Fast-dLLM} baseline, we explicitly disable parallel decoding. This forces all systems to execute the same number of denoising steps, ensuring that latency measurements reflect pure per-token processing overhead rather than speculative execution gains.
    \item \textbf{Hardware Saturation:} For baselines lacking dynamic scheduling, we conduct preliminary profiling to empirically determine the maximum static batch size that fits in memory. For \sysname, we configure the token and logit limits (see Table~\ref{tab:system_config}) to fully utilize the available VRAM on the RTX 4090 and L40S respectively, preventing Out-Of-Memory (OOM) errors while maximizing concurrency.
\end{itemize}

\mytitle{Workloads} We use several workloads for different tasks.
\begin{itemize}[leftmargin=*]
    \item \textbf{LiveBench} is a benchmark for LLMs designed with test set contamination and objective evaluation in mind. We select LiveBench Coding section in the experiment.
    \item \textbf{The OpenAI Summarization Comparison} (OSC) dataset is a publicly available resource comprising input texts, each paired with a human-selected ('chosen') summary and a 'rejected' alternative, all originating from real-world interactions between humans and a chatbot.
    \item \textbf{BurstGPT} (Burst) is a ChatGPT(GPT-3.5) \& GPT-4 Workload Trace to Optimize LLM Serving Systems.
    \item \textbf{Grade School Math 8K} (GSM8K)  is a dataset of 8.5K high quality linguistically diverse grade school math word problems. The dataset was created to support the task of question answering on basic mathematical problems that require multi-step reasoning.
    \item \textbf{HumanEval} dataset released by OpenAI includes 164 programming problems with a function sig- nature, docstring, body, and several unit tests. They were handwritten to ensure not to be included in the training set of code generation models.
    
\end{itemize}

\begin{figure*}[htbp]
    \centering
    \includegraphics[width=0.8\textwidth]{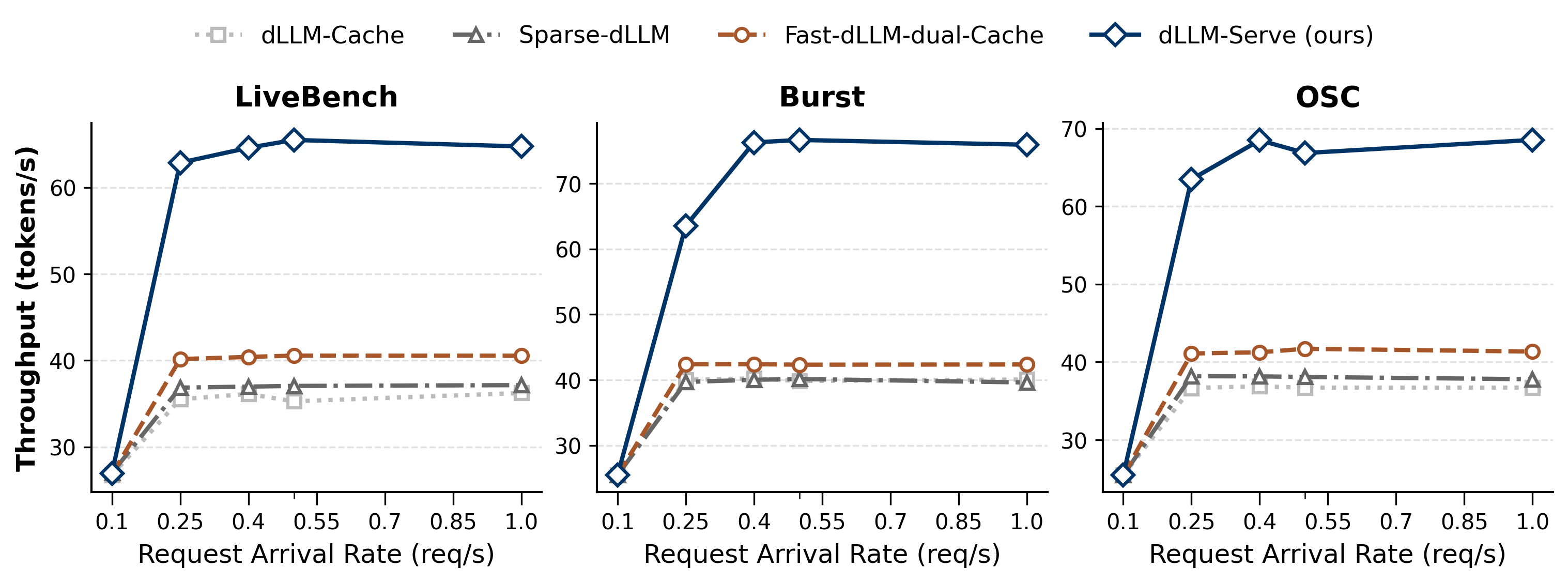}
    \caption{Throughput scalability under increasing arrival rates on the RTX 4090. The plot compares the realized serving throughput (Y-axis) against the request arrival rate (X-axis) for LLaDA-8B.} 
    \label{fig:tp_4090}
\end{figure*}
\subsection{Serving Throughput \& Scalability} \label{sec:eval:throughput}

We evaluate the end-to-end serving capacity of \sysname against three state-of-the-art dLLM serving baselines: \emph{dLLM-Cache}, \emph{Fast-dLLM} (Dual-Cache mode), and \emph{Sparse-dLLM}. Figure~\ref{fig:tp_4090} illustrates the system throughput on the consumer-grade RTX 4090, while Table~\ref{tab:l40s_1qps_results} extends this evaluation to the server-grade NVIDIA L40S.

\mytitle{Peak Capacity Comparison} Across both hardware architectures, \sysname consistently delivers the highest saturation throughput. On the RTX 4090 (Figure~\ref{fig:tp_4090}), \sysname reaches a peak of 76.6 tok/s on the Burst dataset, representing a $1.81\times$ speedup over the strongest baseline (\emph{Fast-dLLM}, 42.4 tok/s). 

This advantage amplifies on server-grade hardware. As shown in Table~\ref{tab:l40s_1qps_results}, running \sysname on the L40S yields a throughput of 106.95 tok/s on the same dataset which is a massive 3.12$\times$ speedup over the baseline. This confirms that our memory-aware budgeting scales effectively with larger VRAM capacities, converting the L40S's extra resources into strictly higher concurrency rather than overhead.

\mytitle{Scalability under Load} The results highlight a critical architectural distinction. As shown in Figure~\ref{fig:tp_4090}, baseline systems exhibit a sharp "throughput wall" around 0.25 RPS. Beyond this arrival rate, their token generation rate flattens, indicating that the GPU is fully locked by the high memory requirements of the "Refresh" phase. \sysname, however, maintains linear scalability up to 0.4--0.5 RPS on the 4090 and extends even further on the L40S. By interleaving the activation-light "Reuse" steps of active requests with the activation-heavy "Refresh" steps of new requests, our Phase-Multiplexed Scheduler effectively delays system saturation.

\begin{table}[t]
\centering
\caption{\textbf{Hardware Generalization (NVIDIA L40S) at 1.0 req/s.} Throughput and latency comparison on server-grade hardware.}
\label{tab:l40s_1qps_results}
\resizebox{\columnwidth}{!}{
\begin{tabular}{l|l|cc|c}
\toprule
\textbf{Dataset} & \textbf{System} & \textbf{Avg Latency} & \textbf{Throughput} & \textbf{Speedup} \\
& (1.0 req/s) & (s) $\downarrow$ & (tok/s) $\uparrow$ & (vs Base) \\
\midrule
\multirow{4}{*}{\textbf{LiveBench}} 
& dLLM-Cache & 419.17 & 32.52 & 1.01$\times$ \\
& Sparse-dLLM & 435.18 & 32.26 & 1.00$\times$ \\
& Fast-dLLM & 241.19 & 56.46 & 1.75$\times$ \\
& \textbf{\sysname (Ours)} & \textbf{218.50} & \textbf{90.17} & \textbf{2.79$\times$} \\
\midrule
\multirow{4}{*}{\textbf{Burst}} 
& dLLM-Cache & 6766.53 & 33.07 & 0.97$\times$ \\
& Sparse-dLLM & 6466.96 & 34.23 & 1.00$\times$ \\
& Fast-dLLM & 3158.38 & 61.26 & 1.79$\times$ \\
& \textbf{\sysname (Ours)} & \textbf{1551.40} & \textbf{106.95} & \textbf{3.12$\times$} \\
\midrule
\multirow{4}{*}{\textbf{OSC}} 
& dLLM-Cache & 6898.20 & 32.62 & 0.99$\times$ \\
& Sparse-dLLM & 6791.42 & 32.87 & 1.00$\times$ \\
& Fast-dLLM & 3408.36 & 58.23 & 1.77$\times$ \\
& \textbf{\sysname (Ours)} & \textbf{1849.79} & \textbf{94.32} & \textbf{2.87$\times$} \\
\bottomrule
\end{tabular}
}
\end{table}

\begin{figure*}[h]
    \centering
    \includegraphics[width=0.8\textwidth]{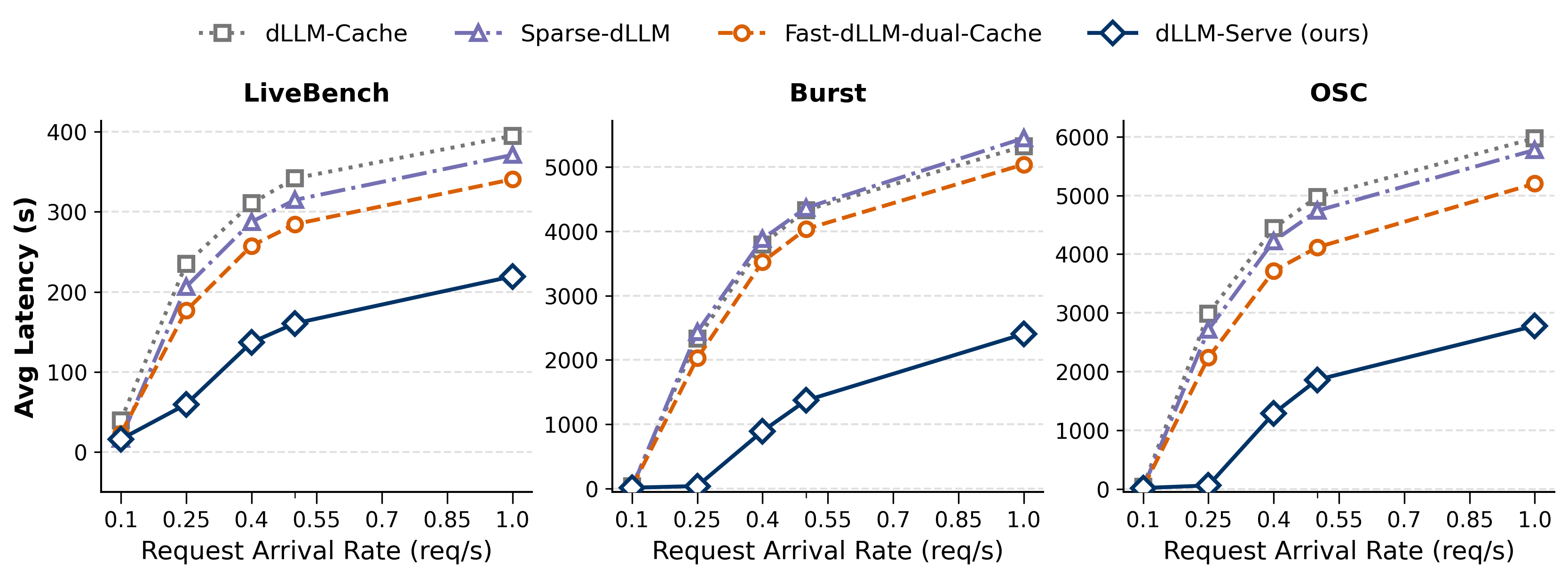}
    \caption{Latency-load comparison of \sysname against three state-of-the-art baselines on the RTX 4090. We report the average end-to-end latency across LiveBench, Burst, and OSC datasets as request arrival rates increase.}
    \label{fig:lat_4090}
\end{figure*}

\subsection{Latency Sensitivity \& Stability} \label{sec:eval:latency}

We further analyze the end-to-end latency of \sysname compared to baselines. Figure~\ref{fig:lat_4090} plots the average query latency as a function of request arrival rate (RPS) on the RTX 4090.

\mytitle{Performance under Contention} At low arrival rates (0.1 RPS), all systems perform comparably, confirming that \sysname incurs negligible scheduling overhead. However, as load increases, the performance gap widens dramatically. On the Burst dataset (RTX 4090), baseline systems suffer catastrophic latency degradation at 0.5 RPS, averaging over 4000s due to resource locking. In contrast, \sysname maintains an average latency of 1370s, a $3\times$ reduction. 
This resilience is even more pronounced on the L40S. As detailed in Table~\ref{tab:l40s_1qps_results}, \sysname reduces the average latency on the OSC dataset from 6791s (Baseline) to just 1849s, a nearly 4$\times$ reduction in wait time. This demonstrates that our budgeting mechanism effectively prevents the "memory gridlock" that paralyzes baseline systems, regardless of the underlying hardware capacity.

\begin{figure}[h]
    \centering
a    \includegraphics[width=0.8\columnwidth]{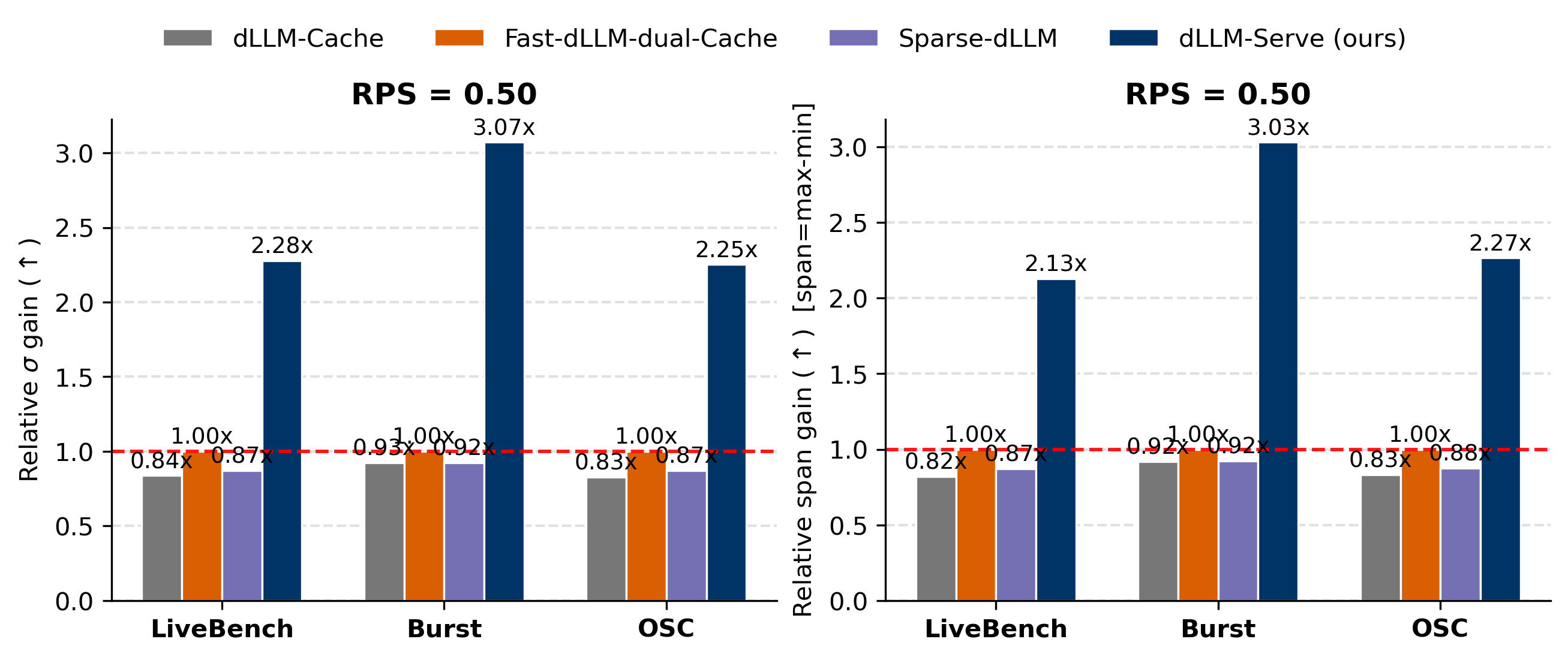}
    \caption{Jitter \& predictability under high load (RTX 4090, RPS=0.5).
    Bars show best-baseline-normalized gains (baseline $=1.0$, red dashed) for latency standard deviation $\sigma$ and tail span $(\max-\min)$ on LiveBench/Burst/OSC; higher is better (less jitter).}
    \label{fig:jitter_predictability_4090}
\end{figure}

\mytitle{Jitter \& Predictability}
Average latency can mask severe tail instability. Figure~\ref{fig:jitter_predictability_4090} reports jitter at high load (RPS=0.5) using latency standard deviation and tail span ($\max-\min$). The baselines exhibit extreme variance: for example, \emph{dLLM-Cache} on Burst ranges from 5.34\,s to 8782.38\,s, which makes interactive serving unpredictable. In contrast, \sysname substantially compresses the tail and improves predictability across all workloads; on LiveBench, it reduces standard deviation by 56\% and tail span by 53\% relative to the best baseline, with similarly large reductions on Burst and OSC. These gains stem from our scheduler, which interleaves Reuse and Refresh requests to avoid head-of-line blocking, preventing short, lightweight requests from being delayed behind long refresh work.

\begin{figure}[h!]
    \centering
    \begin{subfigure}[b]{0.48\columnwidth}
        \centering
        \includegraphics[width=\linewidth]{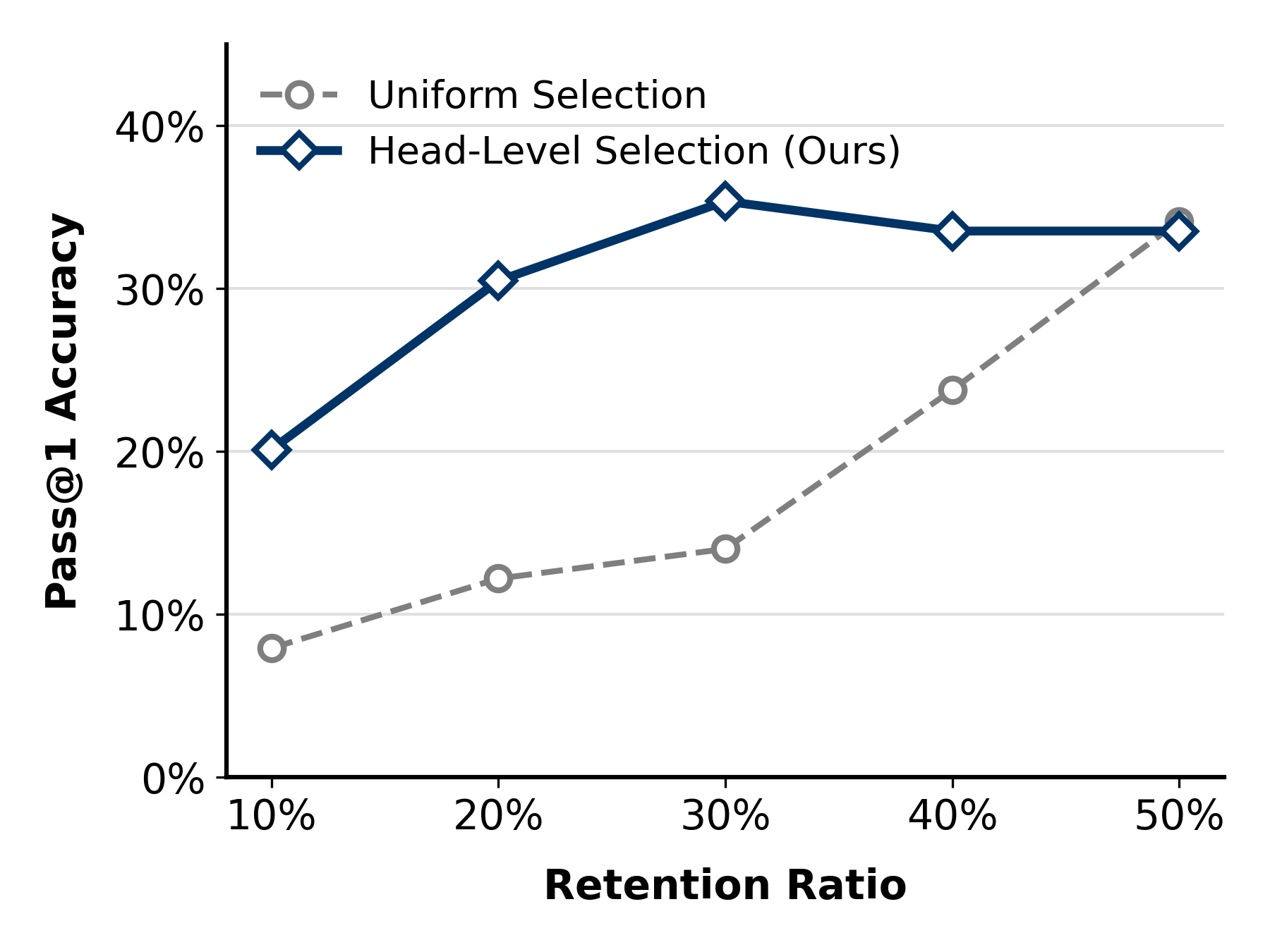}
        \caption{\textbf{HumanEval Pass@1 Accuracy}}
        \label{fig:humaneval_quality}
    \end{subfigure}
    \hfill 
    \begin{subfigure}[b]{0.48\columnwidth}
        \centering
        \includegraphics[width=\linewidth]{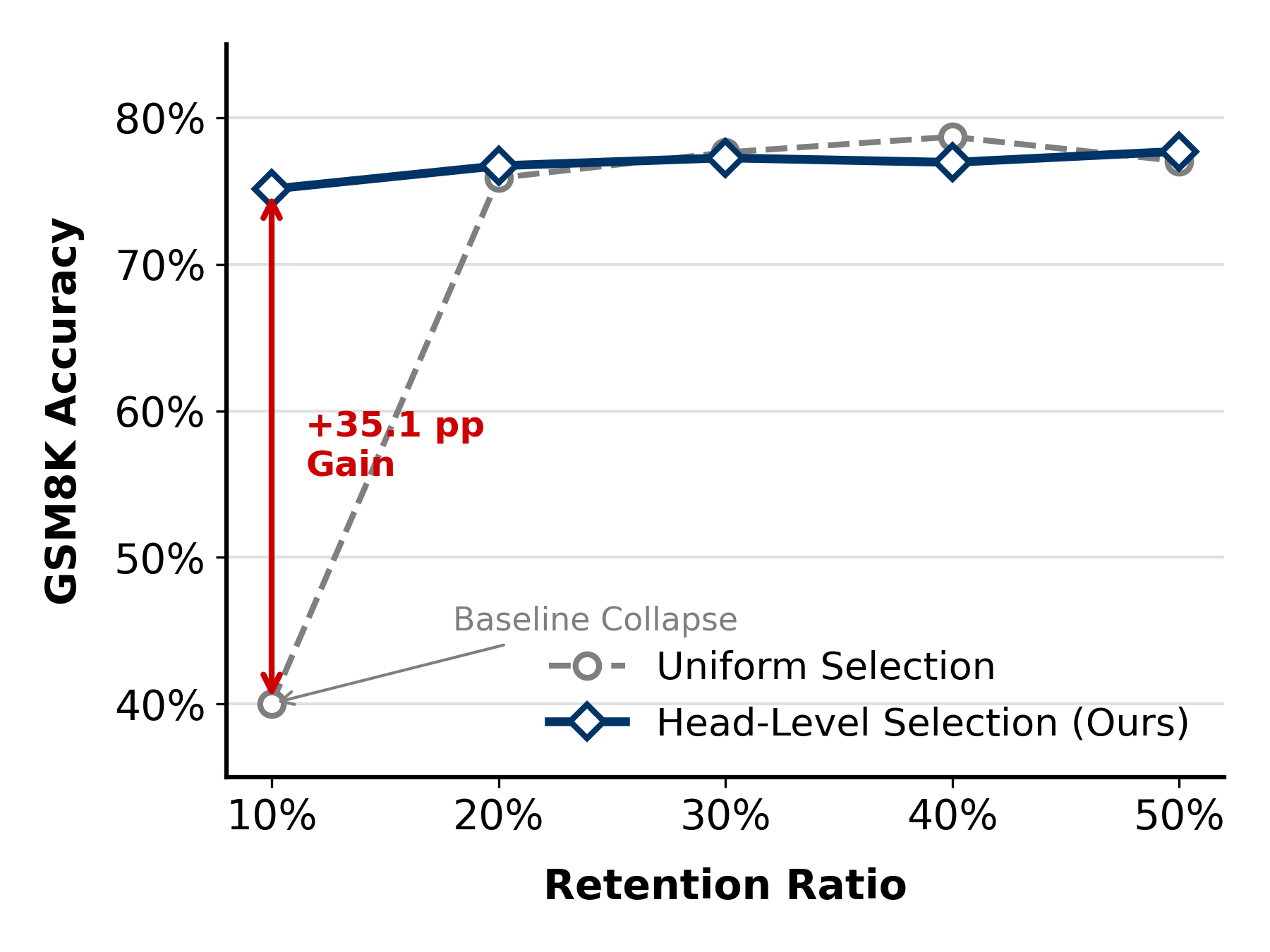}
        \caption{\textbf{GSM8K Accuracy}}
        \label{fig:gsm8k_quality}
    \end{subfigure}
    
    \caption{Accuracy of Head-level Selection (\sysname) vs. Uniform Selection (Sparse-dLLM) under Different Retention Ratios on (a) HumanEval and (b) GSM8K.}
    \label{fig:combined_quality}
\end{figure}



\subsection{Generation Quality Analysis}
\label{sec:eval:quality}
We evaluate whether \sysname's \textit{Head-Centric} sparsity yields higher generation quality than the baseline \textit{Uniform} selection (used in Sparse-dLLM). We test on HumanEval (coding) and GSM8K (math) across retention ratios $r \in \{10\%, \dots, 50\%\}$.

\mytitle{Superiority at Low Retention}
Our Head-Centric approach significantly outperforms the uniform baseline under strict memory constraints. As shown in Figure~\ref{fig:humaneval_quality}, the baseline's coding ability collapses at $r=10\%$ (7.9\% pass@1), while \sysname maintains a usable 20.1\%, a \textbf{$2.5\times$} improvement. This advantage is even more pronounced on GSM8K (Figure~\ref{fig:gsm8k_quality}). At $10\%$ retention, the baseline accuracy drops to 40.0\% (Flexible Extract), whereas \sysname sustains 75.1\%. This \textbf{+87.7\%} relative gain confirms that independent head selection preserves critical semantic features that uniform masks destroy.

Crucially, these results imply that \sysname can achieve target accuracy levels with a significantly smaller memory footprint. For instance, on GSM8K, our method at $r=10\%$ matches the accuracy of the Uniform baseline at $r=20\%$. By halving the required cache size per request without sacrificing quality, \sysname allows the serving engine to fit \textbf{$2\times$} more concurrent requests into the same fixed GPU memory budget. This converts algorithmic precision directly into system-level serving capacity.

\begin{figure}[htbp]
    \centering
    \begin{subfigure}[b]{0.48\columnwidth}
        \centering
        \includegraphics[width=\linewidth]{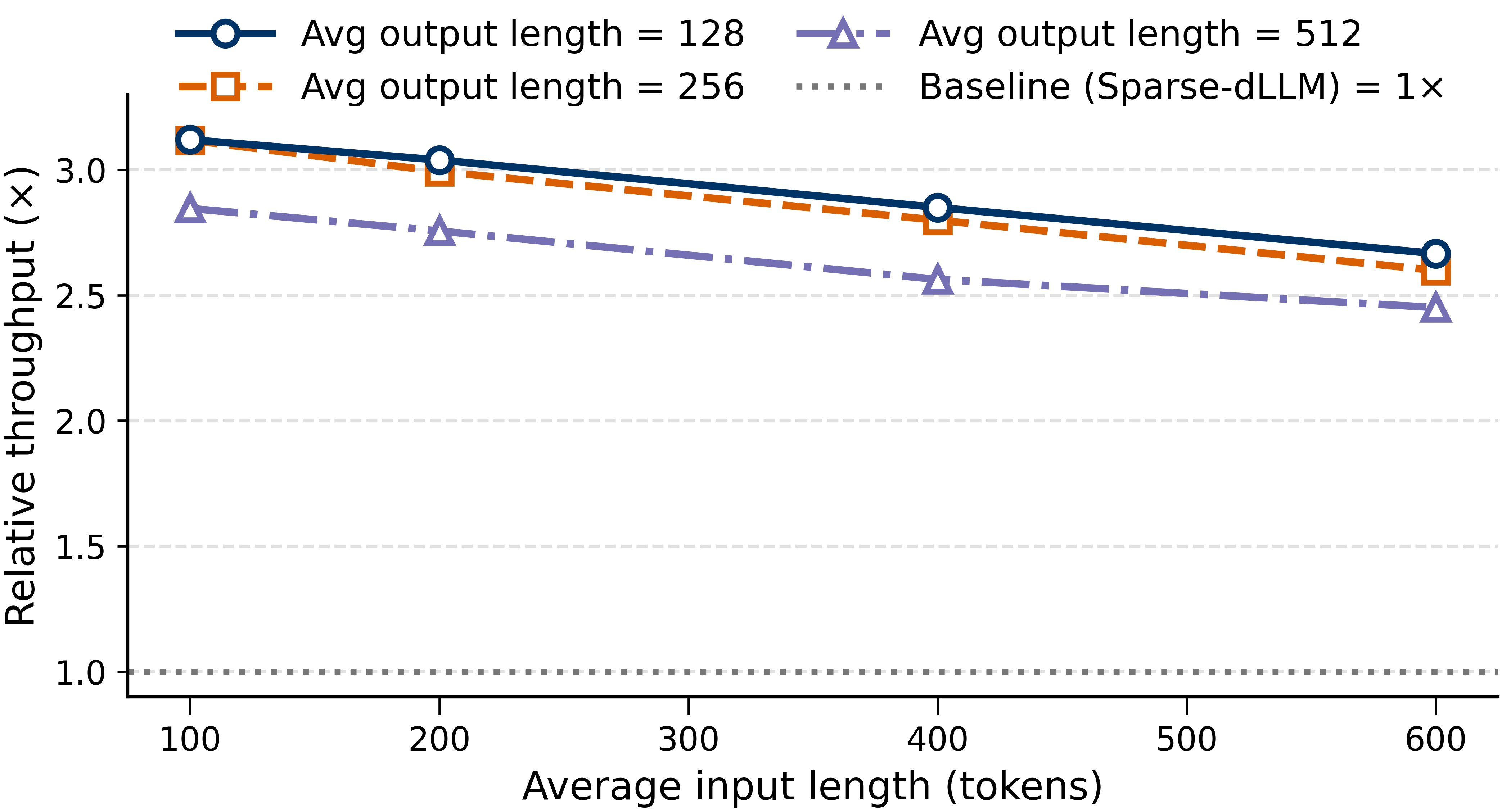}
        \caption{\textbf{Input Length Sweep}}
        \label{fig:input_sweep}
    \end{subfigure}
    \hfill 
    \begin{subfigure}[b]{0.48\columnwidth}
        \centering
        \includegraphics[width=\linewidth]{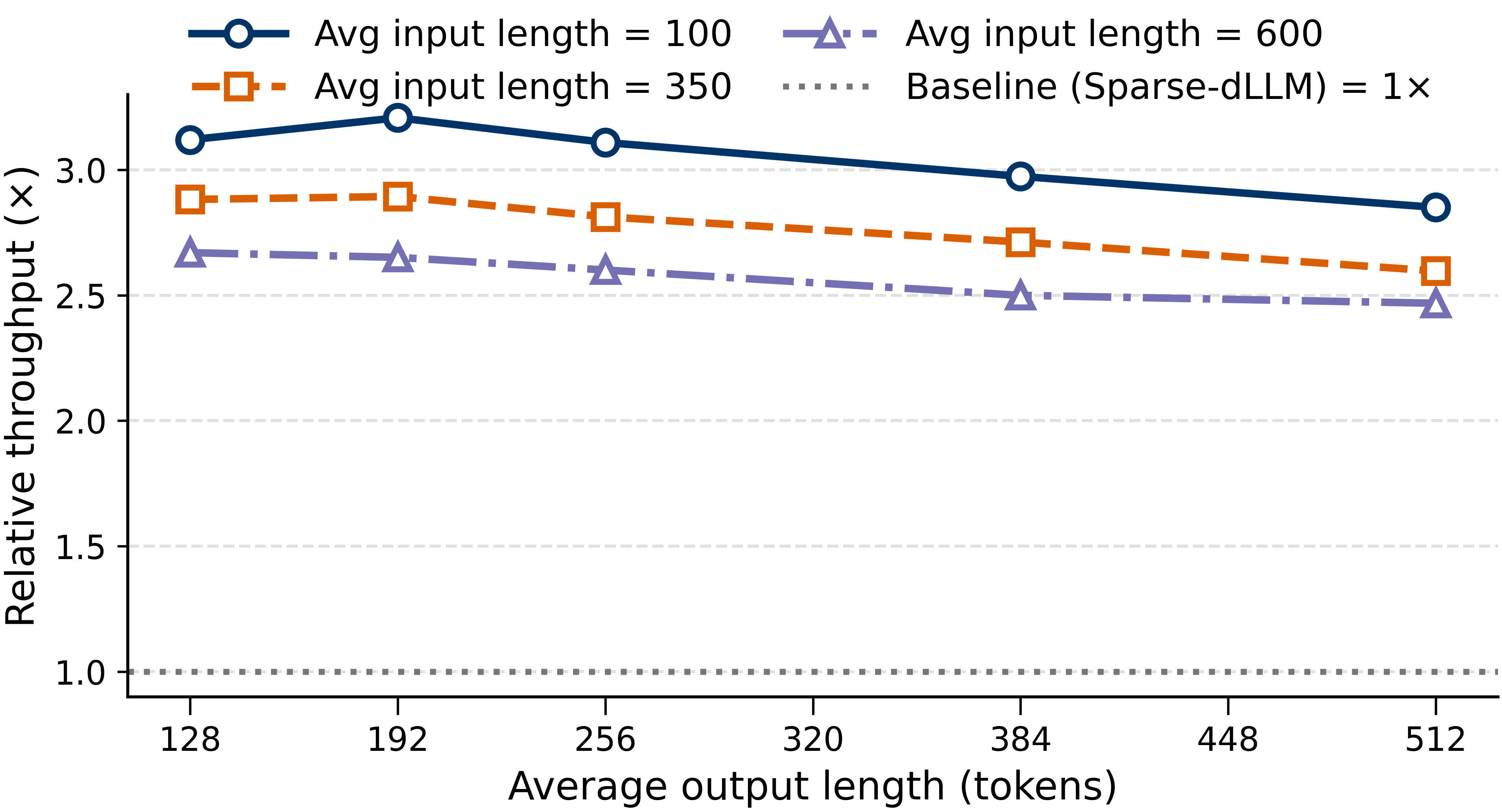}
        \caption{\textbf{Output Length Sweep}}
        \label{fig:output_sweep}
    \end{subfigure}
    
    \caption{The relative throughput of \sysname compared to the baseline (Sparse-dLLM) as a function of (a) average input length and (b) average output length (tokens).}
    \label{fig:combined_sweeps}
\end{figure}



\subsection{Sensitivity Analysis}
\label{sec:sensitivity}

We evaluate the robustness of \sysname by measuring system throughput relative to the \textit{Sparse-dLLM} baseline under varying sequence characteristics. Since \sysname integrates the core sparsity algorithm of \textit{Sparse-dLLM}, we select the standalone \textit{Sparse-dLLM} implementation as our baseline to strictly isolate the performance gains contributed by our system-level designs.

\mytitle{Varying Input Length} As illustrated in Figure~\ref{fig:input_sweep}, we sweep input lengths from $100$ to $600$ tokens while fixing the output size. The data reveals an inverse correlation between prompt length and relative speedup, which decreases from a peak of $3.12\times$ at short contexts to approximately $2.45\times$ at $600$ tokens. This trend occurs because longer prompts increase the duration and memory footprint of the atomic Refresh phase, making these heavy operations harder to effectively hide or interleave with ongoing tasks. 

\mytitle{Varying Output Length} Figure~\ref{fig:output_sweep} plots the performance across output lengths ranging from $128$ to $512$ tokens. We observe a similar stabilization pattern, where speedups reach $3.21\times$ for short generations and settle around $2.47\times$ for longer targets. This reduction indicates that while \sysname excels at managing memory concurrency, extended decoding sequences eventually saturate the GPU's compute capacity, shifting the workload from being memory-limited to compute-bound and slightly narrowing the performance gap.


\subsection{Ablation Study}
\label{sec:ablation}

\begin{figure}[htbp]
    \centering
    \includegraphics[width=\columnwidth]{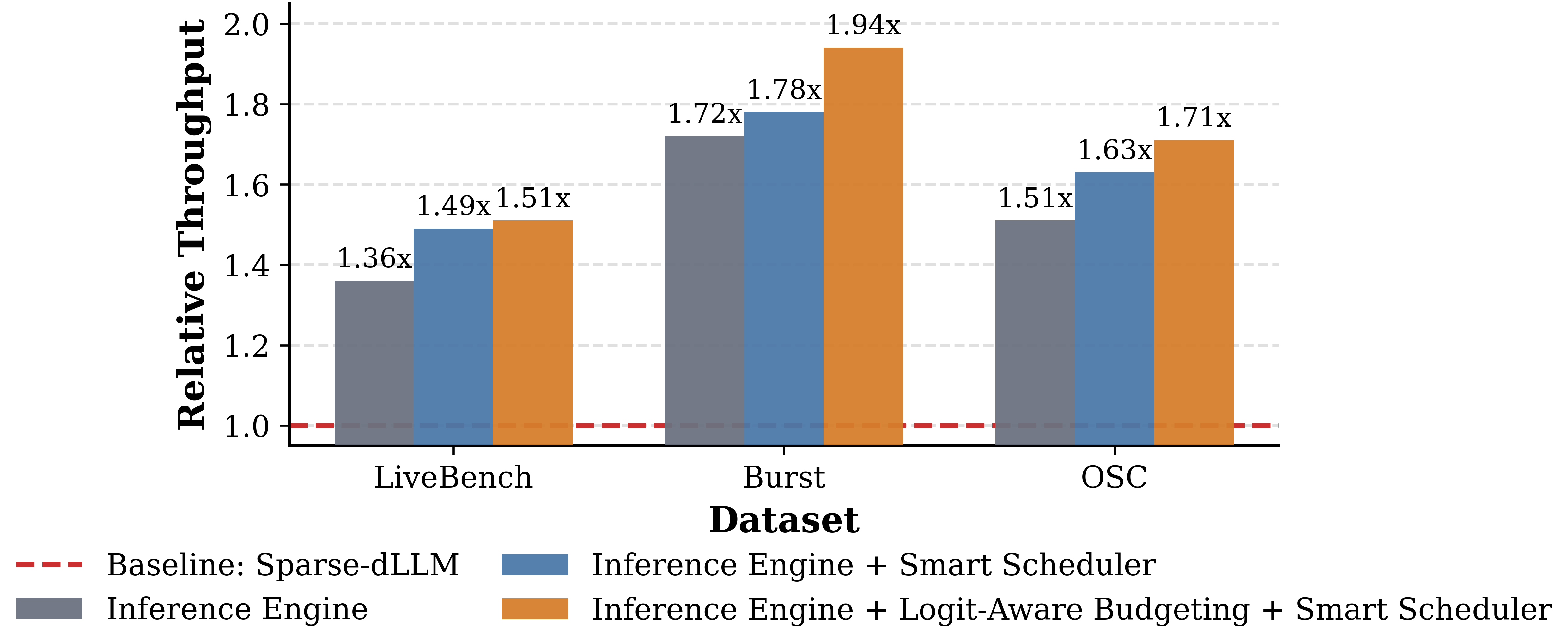}
    \caption{Ablation study of relative throughput on varying datasets.}
\label{fig:ablation_throughput}
    \label{fig:ablation}
\end{figure}

We conduct an ablation study to decouple the performance gains contributed by our three system-level optimizations: the custom Inference Engine, the Phase-Multiplexed Greedy Scheduler (\textit{Smart Scheduler}), and Logit-Aware Activation Budgeting. We treat the standalone \textit{Sparse-dLLM} as our baseline. Figure~\ref{fig:ablation} illustrates the incremental impact of enabling each component.

\mytitle{Inference Engine} We first introduce our custom Inference Engine on top of the baseline. This module integrates FlashAttention, continuous batching (flattening inputs to eliminate padding), and our specialized head-centric KV cache management. As shown in Figure~\ref{fig:ablation}, this architectural overhaul alone delivers the most significant jump in performance, driving throughput speedups of $1.48\times$ on LiveBench, $1.58\times$ on OSC, and $1.76\times$ on Burst datasets. This confirms that optimized memory layouts and cache management are foundational to overcoming the overhead of sparse models.

\mytitle{Smart Scheduler} Next, we enable the Smart Scheduler to optimize request ordering and batch composition. By minimizing pipeline bubbles caused by varying sequence lengths, the scheduler provides an incremental gain, pushing the cumulative speedup to $1.68\times$ on OSC and up to $1.82\times$ on the Burst workload. The impact is most pronounced in these dynamic scenarios, demonstrating the scheduler's ability to effectively handle fluctuating request patterns and maximize resource utilization.

\mytitle{Logit-Aware Budgeting} Finally, we incorporate Logit-Aware Budgeting. This mechanism restricts the memory footprint of the final vocabulary computation to a fixed budget. By preventing the large output tensors from consuming excessive GPU memory, we liberate significant capacity for the KV cache, enabling larger batch sizes and higher concurrency. As evidenced by the full system performance, this reallocation strategy provides the final efficiency boost, elevating the total speedup to $1.75\times$ on LiveBench and $1.76\times$ on OSC, while nearly doubling the baseline performance with a $1.97\times$ speedup on the Burst dataset.
\section{Discussion}
\label{sec:discussion}

\mytitle{Broader Model Support \& Scalability}
We designed \sysname with the primary objective of maximizing serving throughput and minimizing average latency for interactive dLLM workloads. While our evaluation utilizes LLaDA-8B, the current state-of-the-art for diffusion-based language modeling, our contributions target fundamental system-level bottlenecks rather than model-specific artifacts. The oscillating memory demands and iterative denoising phases addressed by our \textit{Logit-Aware Budgeting} and \textit{Phase-Multiplexed Scheduler} are intrinsic to the diffusion generation paradigm, independent of model size. Consequently, the efficiency gains demonstrated here are expected to scale seamlessly to future, larger dLLM architectures. Furthermore, \sysname is architecturally ready for distributed inference: as models scale beyond single-GPU capacities, our \textit{Head-Centric KV Cache} naturally aligns with Tensor Parallelism (TP) sharding strategies, requiring minimal adaptation to coordinate local KV shards across distributed devices.

\mytitle{Budgeting-Aware Scheduling}
Currently, \sysname employs a greedy First-Come-First-Served (FCFS) scheduler. While efficient, this approach optimizes solely for the current iteration's capacity. Due to the distinct memory footprints of the "Refresh" (heavy) and "Reuse" (light) phases, a greedy admission policy can lead to future resource conflicts: a batch that fits in memory at step $t$ may exceed the token budget at step $t+k$ as multiple requests simultaneously enter their refresh phase, forcing delays. Future work will explore a \textit{Budgeting-Aware Scheduler} that incorporates look-ahead planning. By pre-computing the trajectory of token accumulation over the entire generation window, such a scheduler could strictly guarantee that admitted requests never violate the global token cap, thereby eliminating preemption and further reducing average latency.
\section{Related Work}

\mytitle{Inference Optimization for Diffusion LLMs}
\label{sec:related:dllm-infer}
Recent work translates dLLM parallelism into practical speedups via cross-step reuse, cache-compatible decoding structure, and commitment policies. Cross-step caches selectively reuse KV states between denoising iterations by using delayed or conditioned refresh, similarity-gated updates, or approximate projections, substantially reducing recomputation~\cite{ma2025dkvcachecachediffusionlanguage,liu2025dllmcacheacceleratingdiffusionlarge,hu2025flashdlmacceleratingdiffusionlanguage}. Hybrid and blockwise decoders reintroduce cache-friendly structure while preserving within-block bidirectionality through hierarchical caches, parallel pipelines, and adaptive block sizing~\cite{wu2025fastdllmv2efficientblockdiffusion,wu2025fastdllmtrainingfreeaccelerationdiffusion,wang2025diffusionllmsfasterthanarinference,lu2025adablockdllmsemanticawarediffusionllm}. Early-commit and verification strategies reduce both the number of steps and the set of active positions, balancing throughput and accuracy~\cite{li2025diffusionlanguagemodelsknow,israel2025acceleratingdiffusionllmsadaptive,christopher2025speculativediffusiondecodingaccelerating}. Systems integrations demonstrate end-to-end gains beyond microbenchmarks, and large dLLMs achieve competitive tokens-per-second on modern accelerators~\cite{ma2025dinferefficientinferenceframework, song2025seeddiffusionlargescalediffusion}.

\mytitle{Sparse Attention and Cache-Efficient dLLM Decoding}
\label{sec:related:sparse}
The iterative, bidirectional nature of dLLMs creates a different memory–compute profile than AR decoding. Each denoising step attends over the entire prefix and generated region and must consider masked positions, so the naive cost of attention grows quadratically in sequence length and is paid repeatedly over the schedule. Practical systems therefore seek to reduce both the set of tokens that participate in attention and the amount of state that is refreshed at each step.

Cross-step reuse forms the first layer of savings: delayed/conditioned caches (dKV-Cache~\cite{ma2025dkvcachecachediffusionlanguage}) and similarity-driven partial refresh (dLLM-Cache~\cite{liu2025dllmcacheacceleratingdiffusionlarge}) exploit the empirical stability of KV states between adjacent steps. Approximate projections (FreeCache~\cite{hu2025flashdlmacceleratingdiffusionlanguage}) stabilize reuse further when combined with guided unmasking. A second layer targets attention structure itself. \textit{SparseD} observes head-wise regularities across steps and applies head-specific sparse patterns late in denoising, preserving accuracy while reducing memory traffic at long context~\cite{wang2025sparsedsparseattentiondiffusion}. Sparse-dLLM~\cite{song2025sparsedllmacceleratingdiffusionllms} concentrates capacity on salient tokens via dynamic cache eviction driven by attention-guided saliency, which lowers both compute and memory footprint as denoising progresses. Complementary mechanisms such as DPad~\cite{chen2025dpadefficientdiffusionlanguage} restrict suffix computation with a sliding window and distance-decay dropout, and mask-aware budgeting refines which prompt tokens remain resident per head and layer under tight memory limits~\cite{huang2025masktokensprophetfinegrained}.

\section{Conclusion}
\label{sec:conclusion}

In this work, we present \sysname, a holistic serving system designed to bridge the gap between theoretical dynamic sparsity and practical inference acceleration. By co-designing a custom inference engine with a phase-multiplexed scheduler and logit-aware budgeting, \sysname effectively mitigates the system-level bottlenecks inherent to sparse computation. Our evaluation across consumer (RTX 4090) and server-grade (NVIDIA L40S) hardware demonstrates exceptional robustness, achieving up to $1.81\times$ higher throughput and nearly 4$\times$ lower latency compared to state-of-the-art baselines. These results confirm that identifying and managing memory contention is as critical as the sparsity algorithm itself, establishing \sysname as a scalable blueprint for efficient LLM deployment.  We
will open source \sysname codebase to encourage more research
on efficient diffusion LLM serving.

\bibliographystyle{ACM-Reference-Format}
\bibliography{sample-base}

\end{document}